\documentclass[twocolumn,letterpaper,aps,prc,superscriptaddress,nofootinbib,showpacs,floatfix]{revtex4-1}
\usepackage{graphicx}
\usepackage{comment}
\usepackage{setspace}
\usepackage{amsmath}
\usepackage{amssymb}
\usepackage[textwidth=16cm,textheight=22cm]{geometry}
\usepackage{color}
\usepackage{indentfirst}
\usepackage{xspace}
\usepackage{hyperref}
\usepackage{verbatim}
\usepackage{epstopdf}

\usepackage{lineno}

\begin{document}
\title{Transverse Momentum Distribution in Heavy Ion 
  Collision using q-Weibull Formalism }

\author{Sadhana~Dash}
\email{sadhana@phy.iitb.ac.in}
\affiliation{Indian Institute of Technology Bombay, Mumbai,
  India-400076}

\author{ D.~P.~Mahapatra }
\affiliation{Utkal University, Department Of Physics, Vani Vihar, Bhubaneswar, Odisha-751004}

\begin{abstract}
We have implemented the Tsallis q-statistics in the Weibull model of
particle production known as the q-Weibull distribution  to describe
the transverse-momentum ($p_{T}$) distribution of the charged
hadrons at mid-rapidity measured at RHIC and LHC energies. The
model describes the data remarkably well for the entire $p_{T}$ range
measured in nucleus-nucleus and nucleon-nucleon collisions.The proposed
distribution is based on the non-extensive Tsallis q-statistics which
replaces the usual thermal equilibrium assumption of hydrodynamical
models.The parameters of the distribution can be related to the various
aspects of complex dynamics associated with such collision process.

\end{abstract}

\maketitle

\section{Introduction}

The study of relativistic heavy ion collisions provides an excellent means to study
the existence of new form of matter at extreme conditions of energy
density and temperature\cite{starwhite,phenixwhite}.The bulk
properties of the system created in such collisions can be studied via
the transverse momentum ($p_{T}$) distribution of the produced
charged particles which reflect the conditions during the kinetic
freeze out together with the integrated effects of the collective
expansion from very initial stages.The bulk properties of the medium created in such collisions is
usually studied through the statistical approach.One of the well
known approaches to the particle production is governed by
hydrodynamics inspired models based on Boltzman-Gibbs (BG) statistics.The models assumes the occurrence of a local thermal equilibrium
and hence has a common fixed temperature $T$ associated with the
hadronizing system.The transverse momentum($p_{T}$) distributions were explained by the
Boltzman-Gibbs Blast Wave (BGBW) hydrodynamical model fits to the data
to obtain a set of bulk parameters at freeze out such as temperature,$T$ and the radial flow velocity ($\beta$) \cite{starwhite, heinz}.The
successful application of such hydrodynamical models in limited $p_{T}$ range
also provided evidence of collective motion in the continuously
expanding system created in heavy ion collisions. Furthermore, the
measurement of large elliptic anisotropy in momentum in non-central
collisions points towards a nearly thermalized and strongly
interacting system.The equilibrium description based on ideal hydrodynamical models where
BG statistics can be applied are generally limited to low $p_{T}$.It is not always a good idea to select or constrain the $p_{T}$
regime so that the model fitting procedure works.

However, hadronizing systems experience hard QCD scattering  processes  
during initial stages, intrinsic initial state fluctuations due to
formation of initial Color Glass Condensate (CGC) formation ,
fluctuations in temperature , initial energy density {\it etc.} which
can be interpreted as dynamical non-equilibrium effects.These
non-equilibrium processes may not be washed away by multi-particle
interactions in the initial QGP (Quark Gluon Plasma) state or in the
later hadronic phase and can manifest more dominantly in spectral
shape at intermediate or high $p_{T}$\cite{wilk,cgc1,cgc2}.

Recent studies on the generation of azimuthal anisotropy and $v_{2}$
mass splitting using multi-phase transport models suggest the
dominant role of anisotropic parton escape mechanism to describe the   
measured azimuthal anisotropy in small system and heavy ion
collisions.The escape mechanism shows that the most of the $v_{2}$ is
developed during the parton cascade stage and the origin of mass
splitting is attributed to hadron re-scatterings \cite{parton1,
 parton2}.This clearly shows that evolution of the system based on
hydrodynamical scenario is not imperative.In the past,Tsallis non-extensive statistics have been used
extensively to understand the particle production and to describe the evolution of $p_{T}$ spectra in $pp$ and heavy ion collisions at
various energies\cite{tsallis,wilk,wilkosada}.The $q$ parameter in the Tsallis statistics
characterize the deviation from the assumed conditions of 
local thermal equilibrium in systems created in such collisions.
In previous studies,Tsallis distribution have been incorporated into
the Blast Wave model (TBW) and have been used successfully to describe the
spectra at RHIC energies\cite{zebo}.The distribution was also applied
to describe the data in $pp$ collisions at LHC energies to observe the
onset of radial flow in smaller systems\cite{zebopp}.However, the observation of finite radial flow in smaller systems
hinting towards collectivity in such systems is still under
investigation and it is highly debatable whether hydrodynamics can be
applied in smaller systems.

Recently,the Weibull model was used to describe successfully the
particle production in $pp$($p\overline{p}$) and $e^{+}e^{-}$ collisions for a broad range of
energies\cite{weib1,weib2}.A Weibull-Glauber approach was 
also developed to understand the  charged particle multiplicity and
transverse energy distribution in heavy ion
collisions\cite{weib3}.This is natural to assume as the individual nucleon-nucleon collisions
inside the nucleus-nucleus collision has an inherent pQCD parton cascade
evolution and fragmentation.The Weibull distribution is usually used to describe natural
processes where fragmentation and sequential branching is one of the
major ingredients of the dynamical evolution\cite{brown1,brown2}.

The aim of the letter is to propose a generalized distribution within
the framework of non-equilibrium q-statistics which would describe the
$p_{T}$  distribution successfully for both small systems ($pp$ and $pA$)
and the system created in heavy ion collisions for all ranges of
measured $p_{T}$ .

The successful description  of Weibull model describing particle
production can be utilized to incorporate the Tsallis q-statistics to
Weibull distribution to obtain a more generalized q-Weibull
distribution.The q-Weibull distribution has been previously applied to
many complex system in different areas of interest\cite{qweib1,qweib2}
In this letter, we use q-Weibull distribution to describe the $p_{T}$
distribution of charged hadrons emitted in $pp$, $pA$ and $AA$ systems for
broad range of energies.The distribution provides a generalized
statistical framework where  both small and large systems can be
compared.Apart from describing the spectral evolution,the physical
interpretation of the model parameters will provide quantitative
insight into the dynamical processes in such collisions and help in
predicting the spectra for future studies.

\section{The q-Weibull Model}

The most common way to incorporate the q-statistic to the Weibull
model is to replace the exponential factors by their equivalent q-exponential\cite{qweib1}.

The two parameter Weibull distribution  is given by the following expression

\begin{equation}
  P(x;\lambda , k ) = \frac{k}{\lambda} \left (
    \frac{x}{\lambda}\right )^{k-1}~ e^{- (\frac{x}{\lambda})^k}
\end{equation}
  
where $\lambda$ and $k$  are the free parameters.
The q-Weibull distribution is given by \cite{qweib1}

\begin{equation}
  P_{q}(x; q,\lambda , k ) = \frac{k}{\lambda} \left
    (\frac{x}{\lambda}\right )^{k-1}~ e_{q}^{- (\frac{x}{\lambda})^k}
\end{equation}  

where 
\begin{equation}
  e_{q}^{- (\frac{x}{\lambda})^k}  =  (1 -(1-q) (\frac{x}{\lambda})^k)^{ (\frac{1}{1-q})}
\end{equation}  

The value of the non-extensive $q$ parameter is related to the deviation of the
system from thermal equilibrium.In general, $q > 1$ value is
attributed to the presence of intrinsic fluctuations in the system
and depends upon the observable being measured.

\section{$p_{T}$ spectrum and q-Weibull Parameters }

It is worthwhile to investigate whether one can describe the $p_{T}$
distribution of charged hadrons in $pp$, $pA$ and $AA$( heavy ion) collisions
for a broad range of available energies using the q-Weibull approach.This would allow the q-Weibull function the most generalized parametrization to
explain the $p_{T}$ distribution.

\begin{figure}
\includegraphics[scale=0.5]{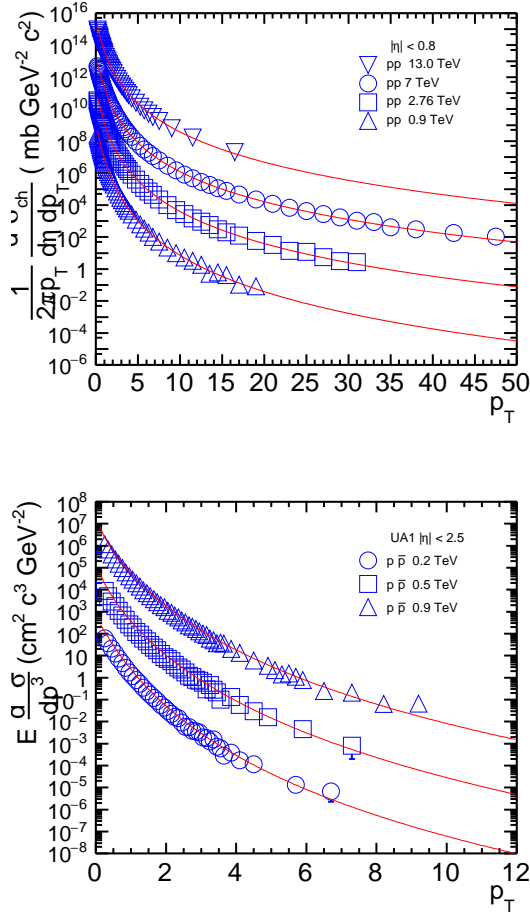}
\caption{(Color online) {\bf (Upper Panel)} $p_{T}$ distribution of charged hadrons in $pp$ 
  collisions at 0.9 TeV, 2.76 TeV  and 7 TeV from pseudo-rapidity region $|\eta| < 0.8$ as measured by ALICE 
  experiment at LHC \cite{alicepp}. 
{\bf (Lower Panel)} $p_{T}$ distribution of charged hadrons in $p\overline{p}$ 
  collisions at 0.2 TeV, 0.5 TeV  and 0.9 TeV in pseudo-rapidity region $|\eta| < 2.5$ as measured by UA1 
  experiment.\cite{ua1}
The solid line is the q-Weibull fit to the data points. The data points are  properly scaled for visibility}
\label{f1}
\end{figure}

\begin{figure}
\includegraphics[scale=0.4]{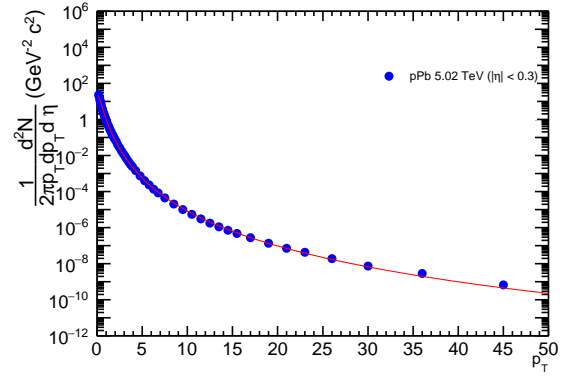}
\caption{(Color online) $p_{T}$ distribution of charged hadrons in $pPb$ 
  collisions at 5.02 TeV  for $|\eta| < 0.3$ as measured by ALICE 
  experiment at LHC \cite{aliceppb}. The solid line is the q-Weibull fit to the data points. The data points are  properly scaled for visibility}
\label{f2}
\end{figure}

\begin{figure}
\includegraphics[scale=0.5]{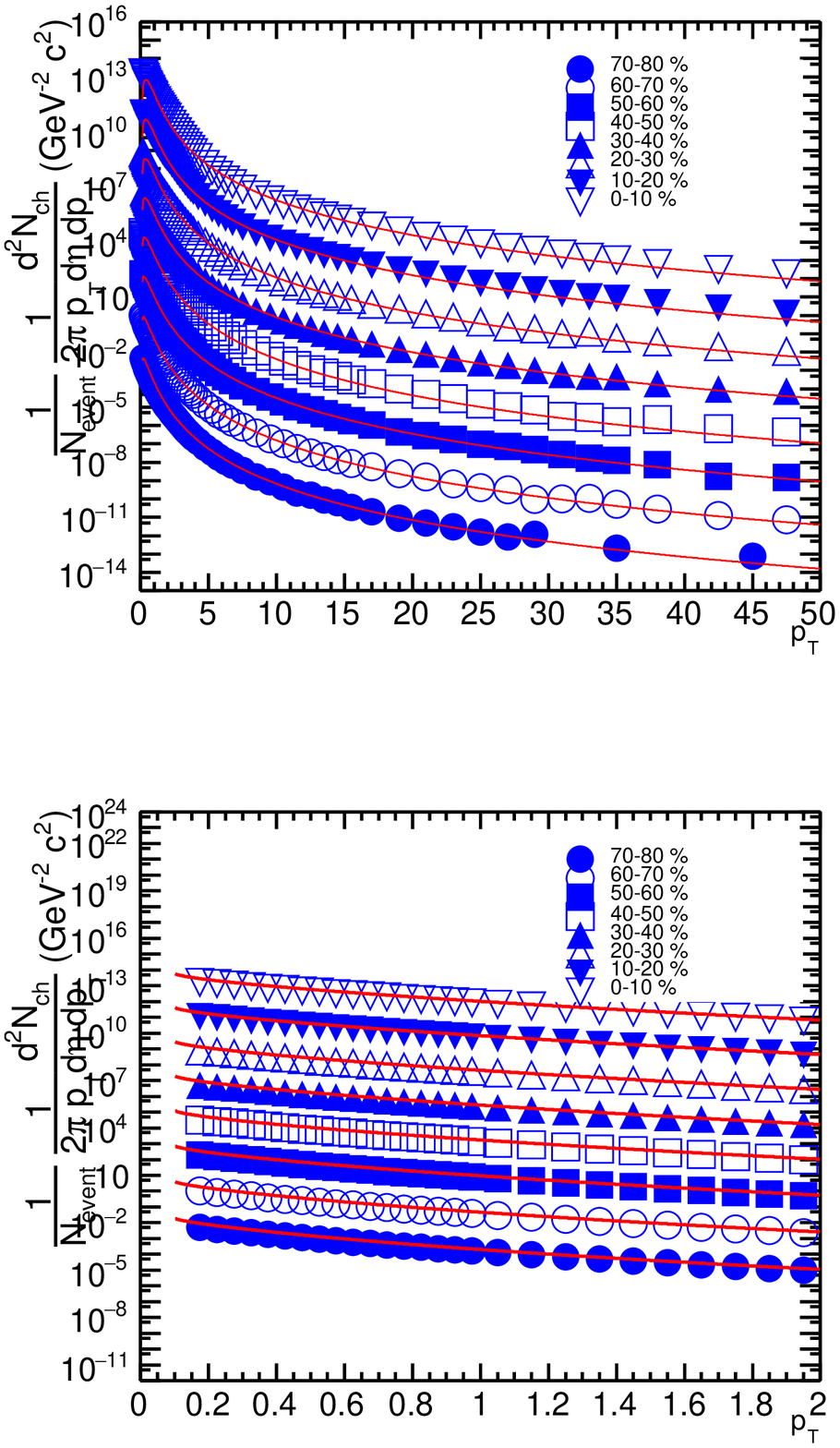}
\caption{(Color online) $p_{T}$ distribution of charged hadrons for various centrality classes  in PbPb 
  collisions at 2.76 TeV from pseudo-rapidity region $|\eta| < 0.8$ as measured by ALICE 
  experiment at LHC \cite{alicepbpb}. The solid line is the q-Weibull
  fit to the data points. The data points are  properly scaled for
  visibility. The upper panel shows the data and fit where the entire
  range of measured $p_{T}$ is taken while the lower panel shoes the
  fit  fot $p_{T} < 2.0 GeV/c$}
\label{f3}
\end{figure}

\begin{figure}
\includegraphics[scale=0.5]{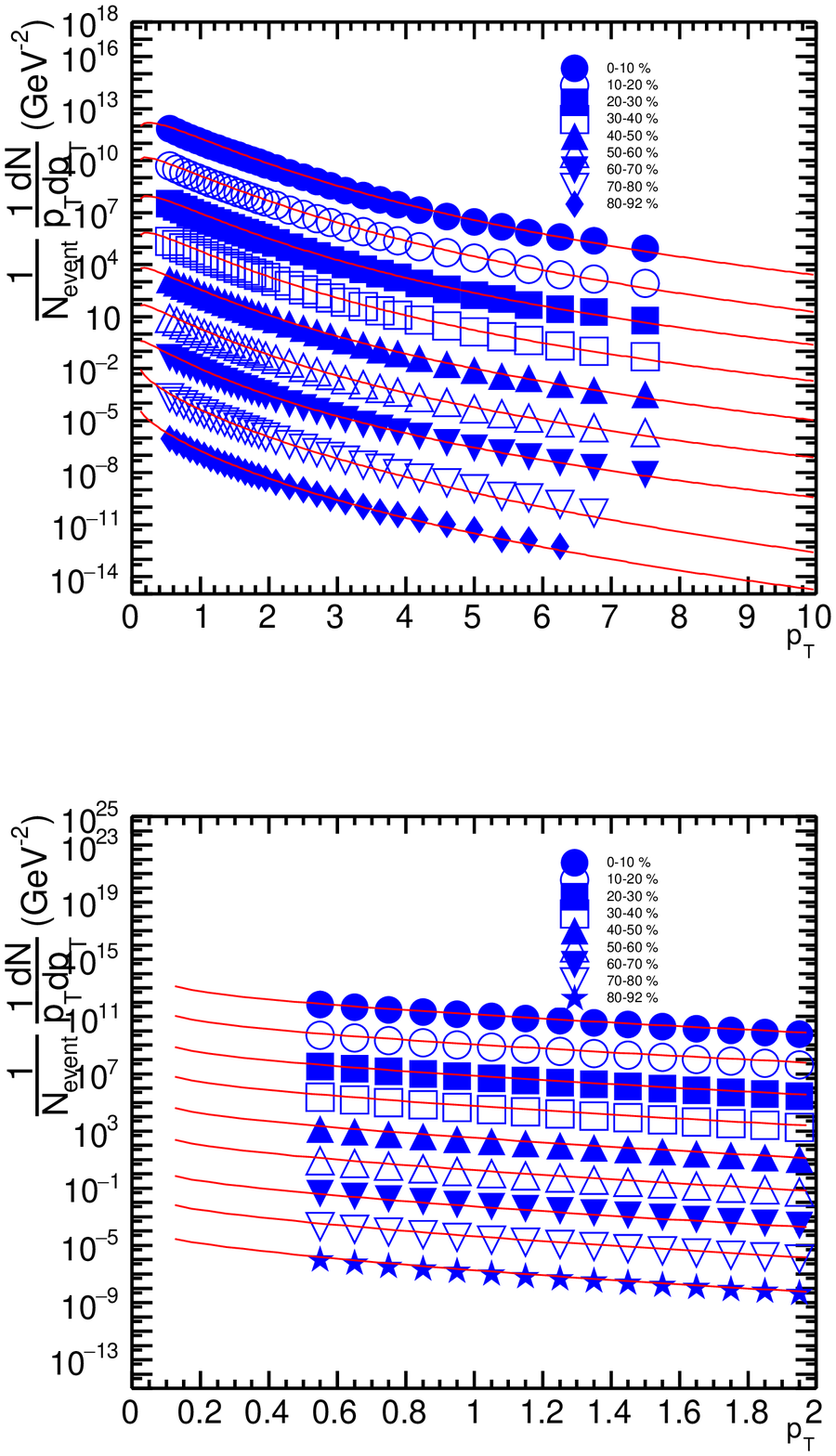}
\caption{(Color online) $p_{T}$ distribution of charged hadrons for various centrality classes  in $AuAu$ 
  collisions at 200 GeV in pseudo-rapidity region $|\eta| < 0.18$ as measured by PHENIX 
  experiment at RHIC \cite{phenix}. The solid line is the q-Weibull
  fit to the data points. The data points are  properly scaled for
  visibility. The upper panel shows the data and fit where the entire
  range of measured $p_{T}$ is taken while the lower panel shoes the
  fit  fot $p_{T} < 2.0 GeV/c$ .}
\label{f4}
\end{figure}

In previous studies \cite{cleyman1, cleyman2}, Tsallis distribution
was successful in explaining the $p_{T}$  distribution of $pp$
collisions for the entire $p_{T}$ range for most of the beam
energies studied. One expects that the q-Weibull function should
describe the $p_{T}$ spectrum for charged particles in $pp$
collisions as it incorporates both the Weibull model of particle
production and the q-statistics which has been very successful so far.The  $p_{T}$ distribution of charged hadrons were fitted with
q-Weibull function in $p\overline{p}$ and $pp$ collisions as measured by UA1 experiment \cite{ua1}
and ALICE experiment at various energies\cite{alicepp}, respectively.This is depicted in Figure~\ref{f1}.The values of the extracted parameters are
shown in Table1 and Table2.It will be interesting to see whether one can extend the
functional description to $pA$ and heavy ion collisions where the
system created has higher degree of complexity and richer features of
particle production mechanism. 
Figure~\ref{f2} shows the description for $pPb$ collisions at 5.02 TeV \cite{aliceppb} . The distribution for $PbPb$
collisions at 2.76 TeV \cite{alicepbpb} and  $AuAu$ \cite{phenix} collisions at 200 GeV for various
centrality classes are shown in Figure~\ref{f3}  and Figure~\ref{f4} respectively.

As can be observed from the figures, the q-Weibull fits provide an
excellent description of the data for different systems and for broad
range of energies in the measured $p_{T}$ range.As q-Weibull appears to be the most generalized q-distribution to
explain the features of  $p_{T}$ distribution, it seems important to study the evolution of its parameters with beam energy, 
type of colliding systems and centrality classes considered.This would allow to attribute some dynamics associated with the collisions processes to the
parameters and thus one can have better understanding of the complex phenomena and have some quantitative estimates for future studies. 

\begin{figure}
\includegraphics[scale=0.4]{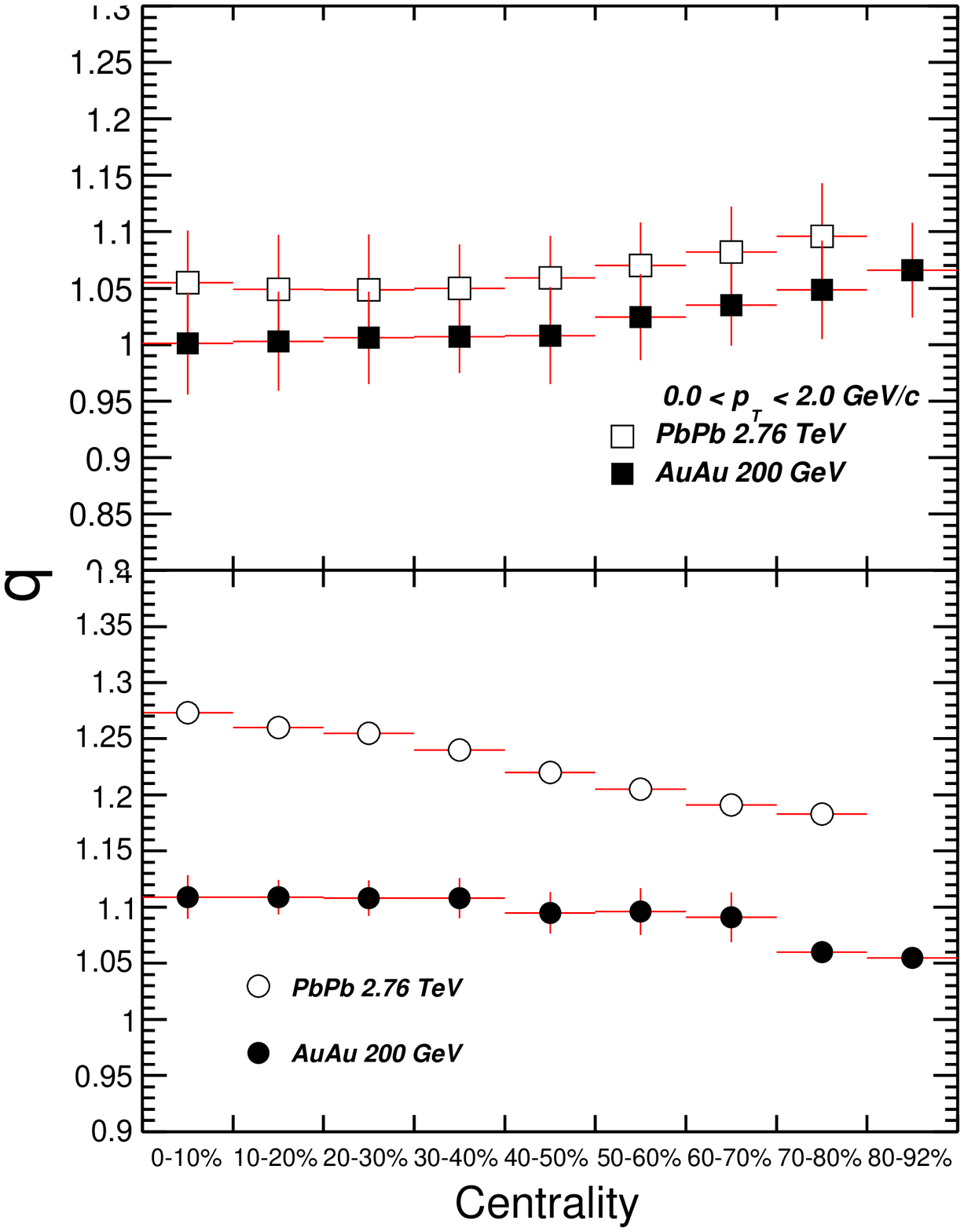}
\caption{(Color online) {\bf (Upper Panel)} Variation of $q$ as a function of 
  centrality for  $p_{T} < 2.0 GeV/c$. {\bf (Lower Panel)} Variation of $q$ as a function of 
  centrality for the entire range of  $p_{T}$ measured. }
\label{f5}
\end{figure}

Figure~\ref{f5} shows the centrality dependence of the  $q$ parameter extracted in $PbPb$ collisions  at 2.76 TeV.  
The non-extensive parameter $q$ which quantifies the  deviation from
local equilibrium goes up as we move from peripheral to central collisions if we consider the entire $p_{T}$ regime for
the fit.The behavior is on expected lines as the initial pQCD hard scattering processes followed by fragmentation and
hadronization dominates the particle production mechanism in hadronic
interactions and supports the non-equilibrium scenario in such
collisions.The processes becomes more 'harder' in more central
collisions and the relative contributions from initial state
fluctuations is also more in central collisions.Hence, we observe an
increasing trend of deviation from equilibrium with centrality. 
However, the $q$ values shows a slight decreasing trend or
remains more or less constant around 1.0 if we constrain the 
$p_{T}$ limit of the fit to lower values ( up to 2 $GeV/c$).This is in agreement with the
hydrodynamical expectation where one sees a local equilibrium to be
attained as the low $p_{T}$ hadrons emanate from a nearly thermally
equilibrated system.The  $q$ parameter in $pp$ collisions is consistently greater than 1.0
and one can see larger $q$ values at highest energies as shown in Table 1.This supports a similar mechanism of  particle production
via initial hard-scatterings across all energies concerned with an increase of relative
contributions of hard or semi hard processes with energy.

\begin{figure}
\includegraphics[scale=0.4]{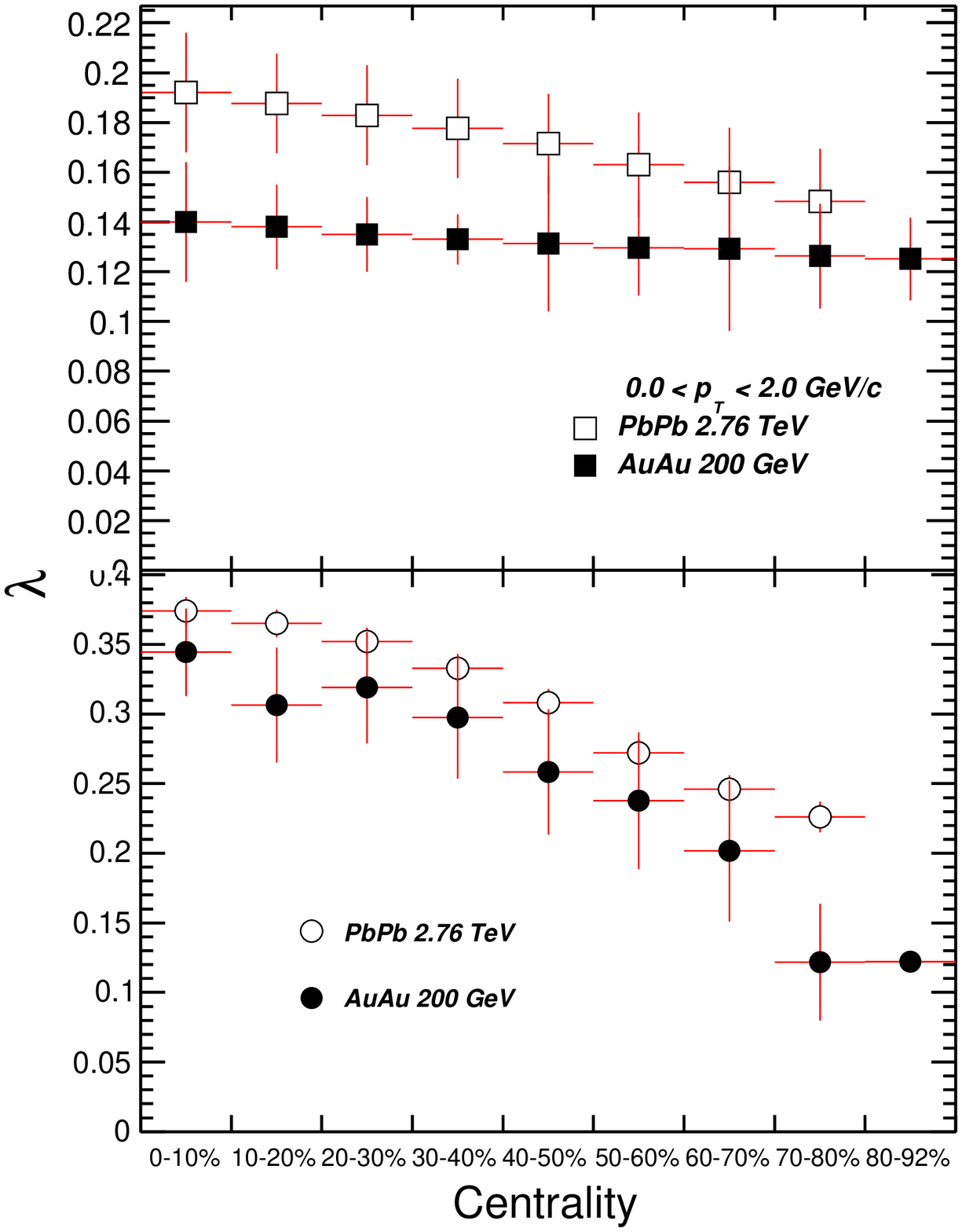}
\caption{(Color online) {\bf (Upper Panel)} Variation of $\lambda$ as a function of 
  centrality for  $p_{T} < 2.0 GeV/c$. {\bf (Lower Panel)} Variation of $\lambda$ as a function of 
  centrality for the entire range of  $p_{T}$ measured. }
\label{f6}
\end{figure}

Figure~\ref{f6} shows the behavior of $\lambda$ parameter for different
centrality in $PbPb$ and $AuAu$ collisions.The $\lambda$ value shows an
increase from peripheral to central collisions for both integrated
$p_{T}$ regime and the low $p_{T}$ regime.This shows that the
$\lambda$  parameter is related to the mean $p_{T}$ of the
distribution.This can also be related to a common radial velocity
(as per transverse expansion scenario in hydrodynamic evolution) in
the low $p_{T}$ regime which goes up with centrality.

The $\lambda$ parameter in $pp$ collisions is more or less constant 
across the beam energies considered if one restricts the fit to a common
value (16 GeV/c in this case) which is expected.

\begin{figure}
\includegraphics[scale=0.4]{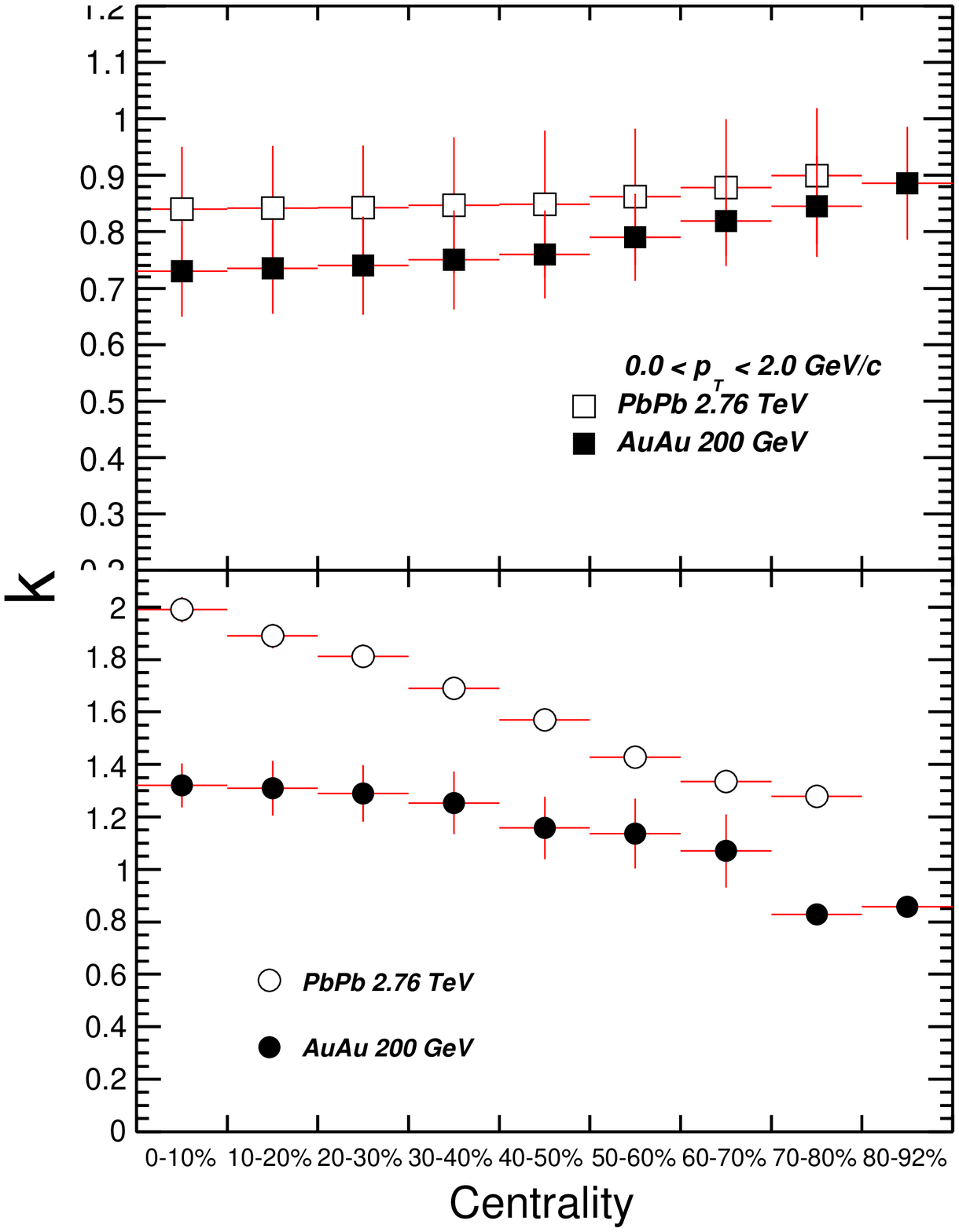}
\caption{(Color online) {\bf (Upper Panel)} Variation of $k$ as a function of 
  centrality for  $p_{T} < 2.0 GeV/c$. {\bf (Lower Panel)} Variation of $k$ as a function of 
  centrality for the entire range of  $p_{T}$ measured. }
\label{f7}
\end{figure}

The $k$ parameter shows a decrease as we move from central to
peripheral collisions in both $PbPb$ and $AuAu$ collisions as shown in
Figure~\ref{f7}. However, the trend is reversed in the low $p_{T}$ regime.
This indicates that $k$ parameter is related to the dynamics of
particle production and its value increases with onset of hard QCD
scatterings, initial fluctuations and other processes leading to  
non-equilibrium conditions.

To understand the physical implications of the parameters better, 
one requires a systematic study of q-Weibull model fits from ISR
to LHC energies in case of hadronic collisions and from SPS to LHC 
energies for heavy ion collisions.The fits to identified particle
spectra will allow deeper understanding of the parameters and 
characterize the system evolution based the species (baryon /meson)  
and mass of the hadrons.The successful description of the spectra and the detailed interpretation of the 
model parameters will certainly be useful in providing qualitative
estimates for future studies.
 
\begin{table}
\centering
\begin{tabular}{|c|c|c|c|}
\hline
$\sqrt{s}$ & $k$           & $\lambda$           & $q$ \\

 TeV &           &           & \\
\hline
0.9  &  1.078 $\pm$ 0.023   & 0.1629 $\pm$ 0.007   & 1.133$\pm$ 0.004 \\
2.76  & 1.09 $\pm$ 0.029       & 0.1711 $\pm$ 0.008   & 1.154$\pm$ 0.005\\
7  & 1.063 $\pm$ 0.03       & 0.1663 $\pm$ 0.008   & 1.1565$\pm$ 0.005\\
13  & 1.010 $\pm$ 0.024      & 0.157 $\pm$ 0.006   & 1.151 $\pm$ 0.005\\
\hline
\hline
\end{tabular}
\caption{ \label{alicefit} The parameters $k$,  $\lambda$ and $q$
  obtained from the fits of $p_{T}$ distributions using q-Weibull
  function in $pp$ collisions ($|\eta| < 0.8$) for different energies as measured by
  ALICE experiment at LHC. The fits are
  constrained  to an upper limit of $p_{T}$ for uniform comparison of parameters.  } 
\end{table}

\begin{table}
\centering
\begin{tabular}{|c|c|c|c|}
\hline
$\sqrt{s}$ & $k$           & $\lambda$           & $q$ \\

 TeV &           &           & \\
\hline
0.2  &  0.91 $\pm$ 0.062   & 0.13 $\pm$ 0.013   & 1.07$\pm$ 0.014 \\
0.5  & 0.921 $\pm$ 0.065       & 0.130 $\pm$ 0.015   & 1.079$\pm$ 0.015\\
0.9  & 0.906 $\pm$ 0.045       & 0.132 $\pm$ 0.019   & 1.080$\pm$ 0.013\\
\hline
\end{tabular}
\caption{ \label{ua1fit} The parameters $k$,  $\lambda$ and $q$
  obtained from the fits of $p_{T}$ distributions using q-Weibull
  function in $p\overline{p}$ collisions ($|\eta| < 2.5$ )for different energies as
  measured by UA1 experiment.  } 
\end{table}

\section{Summary}

We have used the Tsallis q- statistics in the Weibull model  of 
particle production and applied it for the first time  to describe the  transverse momentum distribution 
of charged hadrons in different colliding system for a broad range of energies.The q-Weibull function successfully describes the $p_{T}$ distribution
for  all ranges of $p_{T}$ measured.The parameter $q$, which characterizes the degree of deviation from thermal equilibrium in a 
system decreases systematically from peripheral to the central collisions
in heavy ion collisions if the $p_{T}$ of the particles considered
are constrained to lower values(less than $2 GeV/c$).This indicates an
evolution from  a non-equilibrated system in peripheral collisions
towards a more thermalized system in central heavy ion collisions.However, the trend is reversed if we consider the
all inclusive $p_{T}$ regime which supports an increase of relative 
contribution of hard pQCD processes in central collisions.The
$\lambda$ parameter can be associated with the mean $p_{T}$  or
collective expansion velocity of hadrons which shows an expected
increase with centrality of collisions.The $k$ parameter can be
related to system dynamics associated with the collisions types and
centrality classes.The quantitative evolution of the $k$ parameter and
deeper physical interpretation requires systematic studies of identified particle
spectra for a broad range of energies.We have successfully demonstrated that q-Weibull is the most generalized
statistical model which can be used to describe and predict the $p_{T}$
distribution of charged particles in different collision systems for
broad range of energies.

\section{Acknowledgement}

Sadhana Dash would like to thank the Department of Science and Technology
(DST), India for supporting the present work.


\begin{thebibliography}{50}
\medskip

\bibitem{starwhite} J. ~Adams {\it et al.}, (STAR Collaboration),
  Nuclear Phys. A{\bf 757},102-183 (2005). 

\bibitem{phenixwhite} K. ~Adcox {\it et al.}, (PHENIX Collaboration),
  Nuclear Phys. A{\bf 757},184-283 (2005). 

\bibitem{heinz} E. ~Schnedermann,  J. ~Sollfrank, and U.~ W. ~Heinz,  Phys. Rev. C{\bf 48}, 2462  (1993).

\bibitem{wilk} G. ~Wilk and Z.~ Wlodarczyk, Phys. Rev. Lett.{\bf
    84},2770 (2000)

\bibitem{cgc1} A.~Dumitru, F.~Gelis,~L.~McLerran and R.~Venugopalan, Nucl.Phys.A {\bf 
    810},91 (2008). 

\bibitem{cgc2} S.~Gavin, L.~McLerran and G.~Moschelli, Phys.Rev.C {\bf 
    79},051902 (2009).

\bibitem{parton1} Liang~He, Terrence~Edmonds, Zi-Wei~Lin, Feng~Liu,
  Denes~Molnar and  Fuqiang~Wang,  Phys. Lett. B {\bf 753}, 506-510 
  (2016). 

\bibitem{parton2} Hanlin~Li, Liang~He, Zi-Wei~Lin, Denes~Molnar,
  Fuqiang~Wang and Wei~Xi,  Phys. Rev. C{\bf 93},
  051901(R) (2016)  

\bibitem{tsallis} C. ~Tsallis,~J.Stat.Phys.{\bf 52}, 479(1988)

\bibitem{wilkosada} T.~Osada and G. ~Wilk, Phys. Rev. C{\bf 77},
  044903 (2008) 

\bibitem{zebo} Zebo ~Tang, Yichun ~Xu, Lijuan ~Ruan, Gene ~van~ Buren,
  Fuqiang~ Wang, and Zhangbu~ Xu, Phys. Rev. C{\bf 79}, 051901(R) 
  (2009).

\bibitem{zebopp} Kun ~Jiang, Yinying ~Zhu, Weitao ~Liu, Hongfang
  ~Chen, Cheng ~Li, Lijuan ~Ruan, Zebo~ Tang, and Zhangbu~ Xu,
  Phys. Rev. C{\bf 91}, no. 2, 024910 (2015).



\bibitem{weib1}~S.~Dash, B.~K.~Nandi and P.~Sett, Phys.Rev.D {\bf 
    93},114022 (2016). 

\bibitem{weib2}~S.~Dash, B.~K.~Nandi and P.~Sett, Phys.Rev.D {\bf 
    94}, 074044 (2016).

\bibitem{weib3}~Nirbhay K. Behera, ~S. ~Dash, ~B.~Naik, ~Basanta
  K. ~Nandi and Tanmay Pani, arXiv:1610.02419v1 (2016).

\bibitem{brown1} W~K.~ Brown and K.~H.~Wohletz, J. Appl . Phys.{\bf 78},
  2758 (1995)  
\bibitem{brown2} W~K.~ Brown, J. Astrophys. Astr. {\bf 10}, 89-112 
  (1989)  

\bibitem{qweib1}~S. ~Picoli ~Jr., R. ~S. ~Mendes, and
  L. ~C. ~Malacarne, Physical A {\bf 324}, 678-688 (2003). 

\bibitem{qweib2}~S. ~Picoli ~Jr., R. ~S. ~Mendes, L. ~C. ~Malacarne
  and~R.~P.~B ~Santos, Brazilian Journal of Phys. {\bf 39}, no.2A (2009).


\bibitem{cleyman1} M~D.~ Azmi and J.~Cleymans, J. Phys. G. {\bf 41},
  065001 (2014)  

\bibitem{cleyman2}  J.~Cleymans, ~G.~I.~Lykasov,A.~S.~Parvan,~A.~S.~Sorin,~O.~V.~Teryaev and ~D.~Worku, Phys. Lett. {\bf B723},351-354 (2013)  

\bibitem{ua1}  C. ~Albajar {\it et al.}, (UA1 Collaboration),
  Nuclear Phys. B {\bf 335}, 261-287(1990). 

\bibitem{alicepp}  B. ~Abelev {\it et al.}, (ALICE Collaboration),
  Eur. Phys. J.C {\bf 73:2662}, (2013). 

\bibitem{aliceppb}  B. ~Abelev {\it et al.}, (ALICE Collaboration),
  Phys. Rev. Lett.{\bf 110}, 082302(2013)

\bibitem{alicepbpb}  B. ~Abelev {\it et al.}, (ALICE Collaboration),
  Phys. Lett. B {\bf 720}, 52-62(2013). 

\bibitem{phenix}  S. ~S. ~Adler {\it et al.}, (PHENIX Collaboration),
  Phys. Rev. C{\bf 69},034910 (2004). 












\end{thebibliography}
\end{document}